\documentclass[10pt,aps,prl,twocolumn,amssymb,showpacs,preprintnumbers,nofootinbib]{revtex4-1}
\usepackage{amsmath,amssymb,graphicx}
\usepackage{enumerate}

\pdfoutput=1 

\newcommand{\nc}{\newcommand}
\nc{\non}{\nonumber}
\nc{\beq}{\begin{equation}}   \nc{\eeq}{\end{equation}}
\nc{\bea}{\begin{eqnarray}}   \nc{\eea}{\end{eqnarray}}
\nc{\baa}{\begin{array}}      \nc{\eaa}{\end{array}}
\nc{\bit}{\begin{itemize}}    \nc{\eit}{\end{itemize}}
\nc{\ben}{\begin{enumerate}}  \nc{\een}{\end{enumerate}}
\nc{\bce}{\begin{center}}     \nc{\ece}{\end{center}}

\begin{document}

\preprint{IZTECH/PHYS-2015-05}

\title{\bf Softly Fine-Tuned Standard Model and the Scale of Inflation}

\vskip 20pt

\author{Beste Korutlu\footnote{bestekorutlu@iyte.edu.tr}}
\affiliation{Department of Physics, \.{I}zmir Institute of Technology\\
Urla, \.{I}zmir, 35430 TURKEY}
\date{\today}

\begin{abstract}
The direct coupling between the Higgs field and the spacetime curvature, if finely tuned, is known to stabilize the
Higgs boson mass. The fine-tuning is soft because the Standard Model (SM) parameters are subject to no fine-tuning
thanks to their independence from the Higgs-curvature coupling. This soft fine-tuning leaves behind a large
vacuum energy $\propto \Lambda_{\rm UV}^4$ which inflates the Universe with a Hubble rate $\propto \Lambda_{\rm UV}$,
$\Lambda_{\rm UV}$ being the SM ultraviolet boundary. This means that the tensor-to-scalar ratio inferred from cosmic microwave
background polarization measurements by BICEP2, Planck and others lead to the determination of $\Lambda_{\rm UV}$. The exit
from the inflationary phase, as usual, is accomplished via decays of the vacuum energy. Here we show that, identification
of $\Lambda_{\rm UV}$ with the inflaton, as a sliding UV scale upon the SM, respects the soft fine-tuning constraint and does
not disrupt the stability of the SM Higgs boson.
\end{abstract}


\maketitle

\section{Introduction}	

The Standard Model (SM) suffers from disastrous power-law divergences of quadratic and quartic order \cite{divergence1,divergence2,divergence3,veltman}. The divergences are conveniently parametrized in terms of an ultraviolet (UV) energy scale $\Lambda_{\rm UV}$ --
the ultimate validity limit of the SM. This scale lies below the gravitational scale $M_{Pl}$, as expected of any sensible field theory. The quadratic divergences destabilize the Higgs sector. Now that a scalar consistent with the Higgs Boson of the SM has been discovered at the Large Hadron Collider (LHC) \cite{higgs-exp1,higgs-exp2,higgs-exp3,Ellis:2013lra1,Ellis:2013lra2}, discerning this problem has become crucial. The quartic divergences, on the other hand, lead to gargantuan vacuum energies. Recently, by Demir \cite{demir}, it has been shown that the one-loop quadratic divergences can be suppressed completely
if Higgs coupling to spacetime curvature is finely tuned. The most interesting aspect of this fine-tuning is that it is phantasmal if gravity is classical. The reason for
this phantom behaviour is that the Higgs-curvature coupling does not appear in quantum corrections to the SM parameters. Moreover, particle masses are sensitive only to the Higgs vacuum
expectation value (VEV); they are completely immune to what mechanism has set the Higgs VEV to that specific value appropriate for electroweak interactions. In this sense, one
is able to stabilize the Higgs boson through a ``soft fine-tuning'' that does not interfere with workings of the SM \cite{demir} (see Refs. \cite{also-demir1,also-demir2} for quantum corrections in curved background).

In the present work, we discuss implications of the quartic divergences. More specifically, we show that the quartic divergences induce an enormous vacuum energy which
can inflate the Universe. The scale of inflation sets the UV scale and determines the degree of soft fine-tuning. In fact, quartic contributions
give the plateau section of the slow-roll inflaton potential and fully governs the inflationary epoch for parameter ranges preferred by the softly fine-tuned Higgs mass. As a
matter of fact, an analysis of the inflationary phase is rather timely since recent measurements of the tensor-to-scalar ratio $r$ in CMB polarization have the potential to
fix the scale of inflation. The Polarbear experiment produced the first direct measurement on the polarization of CMB \cite{Ade:2013gez1,Ade:2013gez2}. Then, the BICEP2 collaboration, claiming the detection of $B$-mode polarization of CMB \cite{Ade:2014xna}, reported  $r=0.2^{+0.07}_{-0.05}$ which
corresponds to a high inflationary scale $H\simeq 10^{16}\ {\rm GeV}$ and favours the simplest single-field model of inflation \cite{Guth:1980zm,linde}. The recent Planck observation also promotes this simplest inflationary picture \cite{Ade:2013zuv}. There have been, however, claims that the BICEP2 results could be dominated by polarized dust \cite{Mortonson:2014bja1,Mortonson:2014bja2},
and claims as such seem to be supported by Planck HFI data \cite{Adam:2014bub} and the most recent Planck results. The measurement of the CMB polarization by different experiments
with increasing precision will enable constraining models of inflation. In this paper, exposed is one such model in which inflationary dynamics and electroweak stability are directly correlated. Extended SM scenarios keeping the Higgs vacuum stable while yielding the high-scale inflation successfully exist in the literature by incorporating either an additional $U(1)_{B-L}$ symmetry \cite{Bhattacharya:2014gva}, with nonminimal coupling of the Higgs kinetic term with the Higgs field \cite{Nakayama:2014koa}, with the Einstein tensor \cite{Germani:2014hqa}, with both the Higgs field and the Einstein tensor \cite{Oda:2014rpa}.

\section{Model}
In Ref. \citep{demir}, one focusses only on matter and forces in the SM plus gravitation, and couples them nonminimally
\bea
\Delta V(H,R)=\zeta\, R\, H^{\dagger}H,
\eea
at the renormalizable level. Herein $R$ is the curvature scalar and $H$ is the Higgs field. This nonminimal interaction leads to the Higgs VEV
\bea
\label{H-bkg}
\upsilon^2=\frac{-m_H^2-\frac{4\,\zeta V_0}{M_{\rm Pl}^2}}{\lambda_H+\frac{\zeta m_H^2}{M_{\rm Pl}^2}},
\eea
and curvature scalar
\bea
\label{R-bkg}
R(\upsilon)=\frac{1}{M_{\rm Pl}^2+\zeta \upsilon^2}\left(4 V_0-\lambda_H \upsilon^4\right),
\eea
where the tree-level Higgs potential
\bea
V(H)=V_0+m_H^2 H^{\dagger}H+\lambda_H (H^{\dagger}H)^2,
\eea
encodes the primordial vacuum energy $V_0$, the Higgs squared-mass parameter $m_H^2$, and the Higgs
quartic coupling $\lambda_H$. The Higgs background (\ref{H-bkg}) and curvature background
(\ref{R-bkg}) are obtained by a self-consistent solution of the Higgs motion equation
and Einstein field equations.

Physically, the Higgs VEV $\upsilon$ must lie far below $\Lambda_{\rm UV}$ for $\Lambda_{\rm UV}$ to serve as the UV scale of
the whole setup. In fact, as the definition of electroweak scale, one has to have
\begin{eqnarray}
\label{v-scale}
\upsilon \sim m_{EW},
\end{eqnarray}
where $m_{EW}\ll \Lambda_{\rm UV}$ is the Fermi scale. We take the quartic coupling of Higgs to be $\lambda_H\sim \mathcal{O}(1)$ and $m_H^2\sim m_{EW}^2$ which yields $|\zeta m_H^2|/M_{\rm Pl}^2\ll\lambda_H$. Thus, one can write the Higgs VEV approximately as
\bea\label{v-app}
\upsilon^2\simeq\frac{-m_H^2-\frac{4\,\zeta V_0}{M_{\rm Pl}^2}}{\lambda_H},
\eea
In TABLE I, we give the allowed parameter space satisfying $\upsilon \sim m_{EW}$.
\begin{table}[h!]
{\begin{tabular}{@{}cccc@{}} \toprule
Parameter Space & $\upsilon^2\simeq$ & Constraint\\
\colrule
  $m_H^2\!<0, |\zeta V_0|\!<0$ & $\frac{|m_H^2|+4\,|\zeta V_0|/M_{\rm Pl}^2}{\lambda_H}$ & $|\zeta V_0|\leq |m_H^2|M_{\rm Pl}^2 $ \\
$m_H^2\!<0, |\zeta V_0|\!>0$ & $\frac{|m_H^2|-4\,|\zeta V_0|/M_{\rm Pl}^2}{\lambda_H}$ & $|\zeta V_0|\ll |m_H^2|M_{\rm Pl}^2 $
\\
$m_H^2\!>0, |\zeta V_0|\!<0$ & $\frac{-|m_H^2|+4\,|\zeta V_0|/M_{\rm Pl}^2}{\lambda_H}$ & $|\zeta V_0|\sim |m_H^2|M_{\rm Pl}^2/2$\\
$m_H^2\!>0, |\zeta V_0|\!>0$ & $\frac{-|m_H^2|-4\,|\zeta V_0|/M_{\rm Pl}^2}{\lambda_H}$ & $\qquad\,\,\,\,\,\,\,\,\,$- \\ \botrule
\end{tabular}
\caption{Constraint on $\zeta V_0$ for different parameter spaces at tree-level.}
\label{table1}}
\end{table}
This condition is fulfilled without excessive fine-tuning by implementing the constraints from the last column of TABLE I for the chosen parameter space. Note that for each case $V_0$ lies at an intermediate scale, far below $M_{Pl}^4$. As a result of
(\ref{v-scale}) and constraints from TABLE I, the curvature scalar is found to lie at the electroweak scale
\begin{eqnarray}
\label{R-scale}
R \sim m_{EW}^2,
\end{eqnarray}
because $M_{Pl}^2 + \zeta v^2 \simeq M_{Pl}^2$. The equations (\ref{v-scale}) and (\ref{R-scale}) ensure that Higgs and curvature
sectors both lie at the Fermi scale for each parameter domain in TABLE I.

\noindent The tree-level results above change after quantum effects are incorporated. As a matter of fact,
the Higgs background  (\ref{H-bkg}) and curvature background (\ref{R-bkg}) receive power-law
quantum corrections which can be way too large to draw near $m_{EW}$. Indeed, as derived
explicitly in Ref \cite{demir}, the Higgs VEV shifts from its phyical value in (\ref{v-scale}) by an amount
\bea
\delta\upsilon^2\simeq \frac{3 Q(\zeta)}{(4\pi)^2\lambda_H}\left(2h_t^2-\frac{3}{4}g_2^2-\frac{1}{4}g_Y^2-2\lambda_H\right)\Lambda_{\rm UV}^2,
\eea
which is nothing but the well-known Veltman correction \cite{veltman} except for the nonminimmallity factor
\bea\label{Qzeta}
Q(\zeta)=1-\frac{\zeta (n_F-n_B)}{3\left(2h_t^2-\frac{3}{4}g_2^2-\frac{1}{4}g_Y^2-2\lambda_H\right)}\frac{\Lambda_{\rm UV}^2}{M_{\rm Pl}^2},
\eea
obtained after neglecting logarithmic UV contributions and dropping minuscule $\mathcal{O}(m_H^2/M_{\rm Pl}^2)$ and $\mathcal{O}(V_0/M_{\rm Pl}^4)$ terms. Here $n_F$ and $n_B$ are the numbers of fermions and bosons in the SM, respectively. This factor, thanks to its explicit $\zeta$ dependence, acts as a new degree of freedom for suppressing the powerlaw divergences without adjusting $m_H^2$, $\lambda_H$,
Yukawa couplings or any other SM parameter. Indeed, one can always set $\zeta$ to a specific value $\zeta = \zeta_{0}$ to make
\bea\label{Qzeta0}
|Q\left(\zeta_0\right)|\leq \frac{\upsilon^2}{\Lambda_{\rm UV}^2},
\eea
so that the desired stable value $\delta\upsilon^2 \leq m_H^2$ is obtained. Solution of $\zeta_0$ from (\ref{v-app}), (\ref{Qzeta}) and (\ref{Qzeta0}), constrains $\zeta_0$ such that
\bea\label{zeta0rest}
\zeta_0\!\!\gtrsim\!\!\frac{3(m_H^2\!\pm\!\lambda_H\Lambda_{\rm UV}^2)\!\left(2h_t^2\!-\!\frac{3}{4}g_2^2\!-\!\frac{1}{4}g_Y^2\!-\!2\lambda_H\!\right)}{\lambda_H(n_F-n_B)}\!\frac{M_{\rm Pl}^2}{\Lambda_{\rm UV}^4}\!,
\eea
where the relative sign is due to the absolute value in (\ref{Qzeta0}), yet the stability of Higgs mass at one-loop excludes the solution of $\zeta_0$ with negative sign. In fact, one can even kill $\delta\upsilon^2$ wholly by setting
\bea
\zeta_0 = \frac{3}{(n_F-n_B)} \left(\!2h_t^2-\frac{3}{4}\!g_2^2-\frac{1}{4}g_Y^2-2\lambda_H\!\right)\! \frac{M_{\rm Pl}^2}{\Lambda_{\rm UV}^2},
\eea
for which $Q\left(\zeta_0\right)$ vanishes. The electroweak stability achieved this way, or less restrictively through (\ref{Qzeta}), rests exclusively on
the fine-tuning of $\zeta$, not on any other parameter in the theory. The SM parameters depend on $\zeta$ neither at tree-level nor at any loop-level thanks
to the classical nature of gravity, and hence, fine-tuning of $\zeta$ does not affect them at all. The only quantity that knows $\zeta$, is the electroweak Higgs
VEV $\upsilon$ whose origin and formation process (by fine-tuning or by another mechanism) does not affect the workings of the SM. This whole mechanism might
be called {\it Soft Fine-Tuning} as it does not touch the SM apart form setting its scale. In TABLE I, we treat $\zeta V_0$ together  concerning the state of being positive or negative, and conclude that the parameter space in the last row $m_H^2\!>0, |\zeta V_0|\!>0$ is not allowed at tree-level. However, when handled individually, for $\zeta>0$, if $V_0<0$ is satisfied, a physical VEV still could be achieved. The only problem is that, $V_0<0$ gives an anti-de sitter spacetime $AdSn$, which is very different from the world we actually live in, yet this complication is easily fixed when loop-effects are taken into account. The quantum correction to vacuum energy is such that $\delta V_0>0$ and $\delta V_0 \gg V_0$, resulting in the familiar de sitter spacetime at loop-level. Therefore, the last raw in TABLE I, despite being unphysical at tree-level, is physical at loop-level.

We are done with the electroweak sector. But, what about the curvature induced by vacuum energy? Indeed, the condition (\ref{Qzeta0}) on $\zeta$, imposed
for suppressing the power-law quantum corrections to the electroweak scale $\upsilon$, causes no suppression for vacuum energy \cite{demir}.
As a matter of fact,
quantum corrections shift the scalar curvature by
\begin{eqnarray}
\label{deltR}
\delta R = \frac{(n_F-n_B)}{(4\pi)^2} \frac{\Lambda_{\rm UV}^4}{M_{Pl}^2},
\end{eqnarray}
as follows from (\ref{R-bkg}) after neglecting subleading quadratic and logarithmic corrections. (Quantum corrections also induce logarithmic quadratic curvature contribution $R^2$ and Weyl contribution $W^2$, which we neglect in the subsequent analysis \cite{demir}.) This curvature correction, proportional to $n_F-n_B$, takes enormous values if there is no fermion-boson degeneracy in the theory. The SM exhibits no such symmetry. In case of fermion-boson symmetry, which
seems not existent at the electroweak scale, scalar curvature stays stabilized at the electroweak scale, $\delta R \propto  \left(\frac{\Lambda_{\rm UV}^2}{M_{Pl}^2}\right) m_H^2$.
Leaving aside this possibility, the background de Sitter spacetime is found to have the Hubble constant ${\mathcal{H}}^2 = \delta R/(12)$ which follows from the definition of the Ricci Scalar
\bea
 R(g)=6\left[\frac{\ddot{a}}{a}+\left(\frac{\dot{a}}{a}\right)^2\right],
\eea
for a flat Universe. This gives numerically
\begin{eqnarray}
\mathcal{H} \simeq 0.18 \frac{\Lambda_{\rm UV}^2}{M_{Pl}},
\end{eqnarray}
for the SM spectrum where the Hubble parameter acts like a moment arm in a see-saw creating a balance between $\Lambda_{\rm UV}$ and
$M_{Pl}$. The CMB observations such as BICEP2 and Planck can measure $\mathcal{H}$ when foreground is small. This then fixes $\Lambda_{\rm UV}$, directly.
For $\mathcal{H} = 10^{16}\ {\rm GeV}$, as reported by BICEP2, one finds $\Lambda_{\rm UV} = 3.7\times 10^{17}\ {\rm GeV}$. This means that
a measurement of the Hubble constant ${\mathcal{H}}$ determines the upper validity limit $\Lambda_{\rm UV}$ of the SM and, in general,
smaller the ${\mathcal{H}}$ smaller the $\Lambda_{\rm UV}$. The ongoing and upcoming experiments on CMB polarization place upper
limits on tensor to scalar ratio $r=\frac{P_T(k)}{P_S(k)}$, where $P_T(k)$ and $P_S(k)$ are the observable power spectra of tensor and scalar perturbations,
respectively (see Refs. \cite{Friedman1,Friedman2} for more details).  The constraint on the $r$-parameter suggests a scale for the inflation rate of the Universe as
$r$ is simply proportional to $\mathcal{H}$. As a result, $\mathcal{H}$ is obtained as a function of the momentum cut-off scale $\Lambda_{\rm UV}$,
the theoretical UV bound of the SM.

In consequence, quartic quantum corrections from matter loops inflate the Universe with a Hubble constant determined by the UV scale $\Lambda_{\rm UV}$. The flatness, homogeneity and isotropy of the observable Universe can be understood by some 60 e-foldings in a rather short time interval. The crucial question concerns exit of the Universe from this exponential expansion phase. This is not possible with a constant vacuum energy. The resolution comes
from the fact that, the vacuum energy does actually change in time due to phase transitions occurring as the Universe expands. In fact, a decaying cosmological constant was proposed decades ago by Dolgov \cite{Dolgov:1982qq}. In this sense, as an inherent assumption in inflationary cosmology, the vacuum energy can be ascribed as the energy density of a slowly-varying real scalar field. It could be modelled in various ways, and slow-roll of the scalar field along the model potential can give a graceful exit from inflationary epoch such that the vacuum energy at the beginning has effectively decayed into matter and radiation during reheating.

One-loop quantum corrections to the parameters in the Higgs potential ($\delta V_0$, $\delta m_H^2$) and to the parameters in gravity sector ($\delta M_{Pl}^2$), after neglecting logarithmic contributions (see Ref. \cite{demir} for explicit expressions of radiative corrections), could be modelled to explain the inflationary scenario, if $\Lambda_{UV}$ is considered to be a sliding scale. The quadratic and quartic divergences, parametrized in terms of $\Lambda_{\rm UV}$, are attributed to be associated with a dynamical scalar $\phi(t)$ --the inflaton field-- or with its mass parameter $m_{\phi}$, such that the initial value of the field is equal to $\Lambda_{\rm UV}$. This model is exclusive in the sense that, at one-loop, the tree-level Higgs mass-squared
\bea
m_h^2=2\lambda_H\upsilon^2,
\eea
remains stabilized, even with sliding cutoff scale and no new interactions are induced between the SM fields and the inflaton field. More specifically, the anticipated interaction vertices of Higgs and inflaton with couplings $\lambda_{h\phi\phi}$ and $\lambda_{hh\phi\phi}$ could have come from the terms $(\delta m_H^2+\zeta\, \delta R)\upsilon\, h$ and $\frac{1}{2}(\delta m_H^2+\zeta\, \delta R)h^2$, respectively, yet they both go to zero as
\bea
\delta m_H^2+\zeta\, \delta R=0,
\eea
for $\zeta=\zeta_0$ (see Ref. \cite{demir} for $\delta m_H^2$ and (\ref{deltR}) for $\delta R$). Furthermore, as the masses of fermions and gauge bosons are protected by chiral and gauge invariances, correspondingly, loop contributions do not induce power-law divergences, and thus, they also do not couple to the inflaton field.
The prescription of the inflationary paradigm arising from sliding scale scenario either accommodates a non-minimal coupling ($\zeta_{\phi}$) of inflaton with Ricci scalar or it is minimally coupled and does not interact with the spacetime curvature.

After identifying $\Lambda_{UV}$ with a scalar field $\phi(t)$ in de Sitter background or with its mass parameter $m_{\phi}$, one arrives at a general scalar-tensor theory
    \begin{equation}
  \!\!\!\!\mathcal{L}\!\!\supset\!\!\!\int\!\!d^4\!x\sqrt{-g}\!\left[\!
    \frac{M_{\rm Pl}^2}{2} R\!-\!\frac{1}{2} g_{\mu\nu} \partial^{\mu} \phi \partial^{\nu} \phi\!-\! \!V(\phi)\!-\!\frac{\zeta_{\phi}}{2} R\phi^2\!\right]\!,\!
    \end{equation}
from which the motion equation for $\phi(t)$ follows to be
    \bea
    \ddot{\phi}+3{\mathcal{H}}\dot{\phi}+\zeta_{\phi}R\phi+\frac{d V(\phi)}{d\phi}=0,
    \eea
where $\zeta_{\phi}$ is the non-minimal coupling of $\phi(t)$. The temporal components of the Einstein equations give the Hubble parameter $\mathcal{H}$
    \begin{equation}
 \!\!\! \!\mathcal{H}\!\!=\!\! \frac{\zeta_{\phi}\phi\dot{\phi}}{M_{\rm Pl}^2-\zeta_{\phi}\phi^2}\!+\!\sqrt{\frac{\zeta_{\phi}^2\phi^2\dot{\phi}^2}{(M_{\rm Pl}^2-\zeta_{\phi}\phi^2)^2}\!+\!\frac{\frac{1}{2}\dot{\phi}^2+V(\phi)}{3(M_{\rm Pl}^2-\zeta_{\phi}\phi^2)}}.
    \end{equation}
It is clear that geometrodynamics involves kinetic term of the scalar, its potential $V(\phi)$ and the varying gravitational constant $M_{Pl}^2 - \zeta_{\phi} \phi^2$. In general, ${\mathcal{H}}$
is positive. Obviously, $\zeta_{\phi} \rightarrow 0$ describes the minimally scalar field dynamics.
The slow-roll conditions
\bea\label{slowroll}
&&\ddot{\phi}\ll H\dot{\phi},\non\\
&&\dot{\phi}\ll H\phi,\non\\
&&\frac{1}{2}\dot{\phi}^2\ll V(\phi),
\eea
are to hold for realizing inflationary phase of the Universe. The recent study \cite{Kallosh:2013tua} found a general condition that one must have $\zeta_{\phi}=100$ to have a universal
attractor. This result, not specific to the present one, implies that the inflaton $\phi(t)$ should couple strongly to curvature scalar for being included in a universal attractor.

Given the regularized action in Ref. \cite{demir}, there arise three possible scenarios to be considered as emerging from sliding cutoff scale:
\begin{enumerate}[(i)]
\item First, assigning $\Lambda_{\rm UV}\rightarrow \tilde{\phi}$ and redefining $\tilde{\phi}(n_F-n_B)^{1/4}\rightarrow \phi$ result in a nonminimally coupled inflaton field with the potential
    \bea
    V(\phi)=\frac{1}{2} m_{\phi}^2 \phi^2+\frac{1}{4}\lambda_{\phi} \phi^4,
    \eea
     %
with
\begin{eqnarray}
m_{\phi}^2=\frac{\left[
    3(2\lambda_H+\frac{1}{4}g_Y^2+\frac{3}{4}g_2^2- 2h_t^2)\upsilon^2+4m_H^2\right]}{(4\pi)^2\sqrt{n_F-n_B}},
\end{eqnarray}
and $\lambda_{\phi}=1/(4\pi)^2$. The non-minimal coupling is obtained to be $\zeta_{\phi}=\frac{1}{6(4\pi)^2\sqrt{n_F-n_B}}$. This scenario, known as chaotic inflation with non-minimal coupling, has been previously studied by Refs. \cite{Futamase:1987ua1,Futamase:1987ua2} and it is found that, slow-roll conditions in (\ref{slowroll}) cannot be realized unless $\zeta_{\phi}\leq 10^{-3}$. As $\zeta_{\phi}\simeq 10^{-4}$ in our scenario, inflationary dynamics could be successfully driven by the inflaton field.
\item Next, redefining the UV cut-off scale as $\Lambda_{\rm UV}^2\rightarrow \tilde{\phi}(t)^2$ yet taking $\Lambda_{\rm UV}^4\rightarrow \tilde{m}_{\phi}^2\tilde{\phi}(t)^2$ gives rise to a non-minimally coupled inflaton field with the potential
    \bea\label{m2phi2}
    V(\phi)=\frac{1}{2}m_{\phi}^2\phi^2,
    \eea
where this time
\begin{eqnarray}
m_{\phi}^2 &=&\frac{1}{2}(n_F-n_B)\tilde{m}_{\phi}^2+4m_H^2\nonumber\\
   &+&3(2\lambda_H+\frac{1}{4}g_Y^2+\frac{3}{4}g_2^2- 2h_t^2)\upsilon^2,
\end{eqnarray}
and $\phi\equiv \tilde{\phi}/(4\pi)$. It is important to note that after this redefinition of $\tilde{\phi}$, the non-minimal coupling of the inflaton with the Ricci scalar assumes the conformal value $\zeta_{\phi}=1/6$ for which the action is invariant under a conformal transformation. In Refs. \cite{Futamase:1987ua1,Futamase:1987ua2} it has been concluded that slow-roll conditions cannot be realized unless $\zeta_{\phi}\ll 1$. This condition is not satisfied in the present scenario, that is, a non-minimally coupled inflaton scenario with $V(\phi)=\frac{1}{2}m_{\phi}^2\phi^2$ fails to realize the inflationary era.
\item Finally, assigning $\Lambda_{\rm UV}^2\rightarrow \tilde{m}_{\phi}^2$, $\Lambda_{\rm UV}^4\rightarrow \tilde{m}_{\phi}^2\tilde{\phi}^2$ and making the redefinition $\phi \equiv \tilde{\phi}/(4\pi)$, a minimally-coupled, radiatively stable scalar field is obtained which has the potential of the form
    \bea
    V(\phi)=\frac{1}{2}m_{\phi}^2\phi^2,
    \eea
where now
\begin{eqnarray}
m_{\phi}^2=\frac{1}{2}(n_F-n_B)\tilde{m}_{\phi}^2.
\end{eqnarray}
In this option, the inflaton field successfully drives the inflation as described by Linde's chaotic inflationary model \cite{linde} such that the inflaton starts at some initial large value where the potential energy dominates the kinetic energy and slowly rolls down the potential well.
\end{enumerate}

\section{Conclusion}
The present work completes the germinal work by Demir\cite{demir} in regard to the role of the vacuum energy. Here we have shown that, quartically divergent vacuum energy induces an enormous vacuum energy at the UV scale, it inflates the Universe as required by observations, and the Universe exits the inflationary phase provided that the vacuum energy decays. This is achieved by identifying the cutoff scale with a real scalar field that acts as inflaton field. We have found that this transcription of the UV scale leads to successfull slow-roll inflationary models without disrupting the soft fine-tuning condition
necessary for stability of the electroweak scale.  In general, cosmological measurements on the Hubble parameter can directly determine the UV scale. In this sense, one ends up with a setup in which UV end of the SM is fixed by the scale of inflation.

\section*{Acknowledgments}

This work has been supported by T\"{U}B\.{I}TAK, The Scientific and Technical Research Council of Turkey, through the grant 2232, Project No: 113C002 and also through the grant 1001, Project No: 115F212. The author wishes to thank D.A. Demir for fruitful discussions, his constructive comments and suggestions. \\



\begin{thebibliography}{0}
%
\bibitem{divergence1}
K.~G.~Wilson,
  Phys.\ Rev.\ D {\bf 3}, 1818 (1971).
\bibitem{divergence2} 
L.~Susskind,
  Phys.\ Rev.\ D {\bf 20}, 2619 (1979).
\bibitem{divergence3}
 C.~F.~Kolda and H.~Murayama,
  JHEP {\bf 0007}, 035 (2000)
  [hep-ph/0003170].
%
\bibitem{veltman}
M.~J.~G.~Veltman,
  Acta Phys.\ Polon.\ B {\bf 12}, 437 (1981).
%
\bibitem{higgs-exp1}
  G.~Aad {\it et al.}  [ATLAS Collaboration],
  Phys.\ Lett.\ B {\bf 716}, 1 (2012)
  [arXiv:1207.7214 [hep-ex]].
\bibitem{higgs-exp2}
  S.~Chatrchyan {\it et al.}  [CMS Collaboration],
  Phys.\ Lett.\ B {\bf 716}, 30 (2012)
  [arXiv:1207.7235 [hep-ex]].
\bibitem{higgs-exp3}
  S.~Chatrchyan {\it et al.}  [CMS Collaboration],
  JHEP {\bf 1306}, 081 (2013)
  [arXiv:1303.4571 [hep-ex]].
%
\bibitem{Ellis:2013lra1}
J.~Ellis and T.~You,
  JHEP {\bf 1306}, 103 (2013)
  [arXiv:1303.3879 [hep-ph]].
\bibitem{Ellis:2013lra2}
A.~Djouadi and G.~Moreau,
  Eur.\ Phys.\ J.\ C {\bf 73}, no. 9, 2512 (2013)[arXiv:1303.6591 [hep-ph]].
%
\bibitem{demir}
  D.~A.~Demir,
  Phys.\ Lett.\ B {\bf 733}, 237 (2014)
  [arXiv:1405.0300 [hep-ph]].
%
\bibitem{also-demir1}
M.~Visser,
  Mod.\ Phys.\ Lett.\ A {\bf 17}, 977 (2002)
  [gr-qc/0204062].
\bibitem{also-demir2}
  D.~A.~Demir,
  arXiv:1207.4584 [hep-ph].
%
\bibitem{Ade:2013gez1}
  P.~A.~R.~Ade {\it et al.}  [POLARBEAR Collaboration],
  Phys.\ Rev.\ Lett.\  {\bf 113}, 021301 (2014)
  [arXiv:1312.6646 [astro-ph.CO]].
\bibitem{Ade:2013gez2}  
  P.~A.~R.~Ade {\it et al.}  [POLARBEAR Collaboration],
  Astrophys.\ J.\  {\bf 794}, no. 2, 171 (2014)
  [arXiv:1403.2369 [astro-ph.CO]].
%
\bibitem{Ade:2014xna}
  P.~A.~R.~Ade {\it et al.}  [BICEP2 Collaboration],
  Phys.\ Rev.\ Lett.\  {\bf 112}, 241101 (2014)
  [arXiv:1403.3985 [astro-ph.CO]].
%
\bibitem{Guth:1980zm}
  A.~H.~Guth,
  Phys.\ Rev.\ D {\bf 23}, 347 (1981).
%
\bibitem{linde}
  A.~D.~Linde,
  Phys.\ Lett.\ B {\bf 108}, 389 (1982).
%
\bibitem{Ade:2013zuv}
  P.~A.~R.~Ade {\it et al.}  [Planck Collaboration],
  Astron.\ Astrophys.\  {\bf 571}, A16 (2014)
  [arXiv:1303.5076 [astro-ph.CO]].
%
\bibitem{Mortonson:2014bja1}
 M.~J.~Mortonson and U.~Seljak,
  JCAP {\bf 1410}, no. 10, 035 (2014)
  [arXiv:1405.5857 [astro-ph.CO]].
\bibitem{Mortonson:2014bja2}
  R.~Flauger, J.~C.~Hill and D.~N.~Spergel,
  JCAP {\bf 1408}, 039 (2014)
  [arXiv:1405.7351 [astro-ph.CO]].
%
\bibitem{Adam:2014bub}
  R.~Adam {\it et al.}  [Planck Collaboration],
  arXiv:1409.5738 [astro-ph.CO].
%
\bibitem{Bhattacharya:2014gva}
  K.~Bhattacharya, J.~Chakrabortty, S.~Das and T.~Mondal,
  JCAP12(2014)001
  [arXiv:1408.3966 [hep-ph]].
%
\bibitem{Nakayama:2014koa}
  K.~Nakayama and F.~Takahashi,
  Phys.\ Lett.\ B {\bf 734}, 96 (2014)
  [arXiv:1403.4132 [hep-ph]].
%
\bibitem{Germani:2014hqa}
  C.~Germani, Y.~Watanabe and N.~Wintergerst,
  JCAP12(2014)009
  [arXiv:1403.5766 [hep-ph]].
%
\bibitem{Oda:2014rpa}
  I.~Oda and T.~Tomoyose,
  Adv.\ Stud.\ Theor.\ Phys.\  {\bf 8}, no. 13, 551 (2014)
  [arXiv:1404.1538 [hep-ph]].
%
\bibitem{Friedman1}
 R.~Easther, B.~R.~Greene, W.~H.~Kinney and G.~Shiu,
  Phys.\ Rev.\ D {\bf 66}, 023518 (2002)
  [hep-th/0204129].
\bibitem{Friedman2}
  R.~Easther and W.~H.~Kinney,
  Phys.\ Rev.\ D {\bf 67}, 043511 (2003)
  [astro-ph/0210345].
%
\bibitem{Dolgov:1982qq}
  A.~D.~Dolgov,
  JETP Lett.\  {\bf 41}, 345 (1985)
  [Pisma Zh.\ Eksp.\ Teor.\ Fiz.\  {\bf 41}, 280 (1985)].
%
\bibitem{Kallosh:2013tua}
  R.~Kallosh, A.~Linde and D.~Roest,
  Phys.\ Rev.\ Lett.\  {\bf 112}, no. 1, 011303 (2014)
  [arXiv:1310.3950 [hep-th]].
%
\bibitem{Futamase:1987ua1}
  T.~Futamase and K.~I.~Maeda,
  Phys.\ Rev.\ D {\bf 39}, 399 (1989).
\bibitem{Futamase:1987ua2}
  E.~Komatsu and T.~Futamase,
  Phys.\ Rev.\ D {\bf 59}, 064029 (1999)
  [astro-ph/9901127].
%
\end{thebibliography}
\end{document}